\documentclass[a4paper,11pt]{article}
\usepackage{jheppub} % for details on the use of the package, please
\usepackage[T1]{fontenc} % if needed
\usepackage{slashed}
\usepackage{diagbox}
\usepackage{makecell}% for lines in a single cell in a table
\usepackage{comment}

\title{\boldmath Testing lepton-flavor-violating decay of doubly charged Higgs bosons in type-II seesaw via photon fusion at the high-energy LHC}

\author[a]{Hang Zhou,}
\author[b,c]{Ning Liu}

\affiliation[a]{School of Microelectronics and Control Engineering, Changzhou University\\Changzhou, 213164, China}
\affiliation[b]{Physics Department and Institute of Theoretical Physics, Nanjing Normal University\\Nanjing, 210023, China}
\affiliation[c]{Nanjing Key Laboratory of Particle Physics and Astrophysics\\ Nanjing, 210023, China}

\emailAdd{zhouhang@cczu.edu.cn}
\emailAdd{liuning@njnu.edu.cn}

\abstract{Tiny neutrino masses can be explained by the type-II seesaw mechanism, where a triplet scalar under $SU(2)_{L}$ is predicted. Collider searches for this exotic scalar have been extensively conducted, especially for its doubly charged component $\Delta^{\pm\pm}$. Utilizing the forward detectors at the Large Hadron Collider (LHC), we study the probing sensitivity for the elastic photon fusion production of the scalars $pp\to p(\gamma\gamma\to\Delta^{++}\Delta^{--})p$ followed by the lepton-flavor-violating (LFV) decay channels $\Delta^{\pm\pm}\to e^{\pm}\mu^{\pm}$. With a high center-of-mass energy of 100 TeV and several luminosity scenarios, we can extensively broaden the exclusion bounds in the parametric space of Br$(\Delta^{\pm\pm}\to e^{\pm}\mu^{\pm})$ versus the triplet scalar mass $m_{\Delta}$. Specifically, at the 100 TeV LHC with an integrated luminosity of 3 ab$^{-1}$, the mass exclusion limit at 95\% C.L. can reach around 1150 GeV with the assumption of inverted neutrino mass hierarchy.}

\begin{document}
\maketitle
\flushbottom

\section{Introduction}
\label{sec:intro}

% paragraph 1: type-ii seesaw
The established phenomena of neutrino oscillations provide unequivocal evidence that neutrinos possess mass, a feature absent in the original formulation of the Standard Model (SM) of particle physics. This discovery necessitates physics beyond the SM to explain the origin and, crucially, the extreme smallness of neutrino masses. A theoretically elegant and deeply studied solution involves the so-called seesaw mechanism, which posits that the observed tiny masses are a consequence of mixing with new, heavy particles. This mechanism is often expressed through the effective Weinberg operator $\mathcal{L}\propto\ell_{L}HH\ell_{L}/\Lambda$~\cite{Weinberg:1979sa}, whose realization at a more fundamental level leads to three canonical types of seesaw models. These types differ in the nature of the new particles introduced: right-handed neutrinos (type-I), an $SU(2)_{L}$ triplet scalar (type-II) and an $SU(2)_{L}$ triplet fermion (type-III)~\cite{minkowski1977,yanagida1979,ms1980,sv1980,mw1980,cl1980,lsw1981,ms1981,flhj1989,ma1998}. This work concentrates on the type-II seesaw framework. Its distinctive feature is the introduction of a scalar field $\Delta$, which transforms as a triplet under the SM gauge group $SU(2)_{L}$. A compelling signature of this model is the prediction of doubly charged scalar components $\Delta^{\pm\pm}$. The generation of neutrino masses in this scenario arises directly from a Yukawa coupling between the triplet and the SM lepton doublets. The required mass suppression emerges naturally after the electroweak symmetry breaking, considering the mixing between the SM Higgs and the triplet scalar $v_{\Delta}\approx\mu v^{2}_{0}/m^{2}_{\Delta}$, with the triplet mass $m_{\Delta}$ being several orders higher than the electroweak scale at $v\approx246$ GeV, where $v_{0}$ and $v_{\Delta}$ are the vacuum expectation values of the SM Higgs and triplet scalars, and satisfy $v^{2}=v^{2}_{0}+v^{2}_{\Delta}\approx246^{2}\,\text{GeV}^{2}$. Meanwhile, the dimensional mixing parameter $\mu$ is possible to be naturally small enough according to 't Hooft naturalness argument~\cite{Senjanovic:1978ev,tHooft:1979rat} so that in scenarios of type-II seesaw, the Yukawa couplings not only relates neutrino oscillation experiments data to collider searches via leptonic decays of the exotic scalars, its mass $m_{\Delta}$ is also allowed to be low enough and potentially testable at accessible energy scales by current collider experiments. Extended upon the minimal type-II seesaw framework, various innovative researches have put forward new ideas connecting the neutrino mass origin with other important issues, including the dark matter candidates and the cosmic baryon asymmetry~\cite{Gu:2009hu,Zhou:2017lrt,Gu:2018kmv,Gu:2023apn}.

% paragraph 2: current search, collider and other exp
Traditionally, the primary avenue for probing these exotic scalars at colliders like the Large Hadron Collider (LHC) is through the Drell-Yan production processes. Mediated by neutral or charged currents, the doubly charged scalars can be produced in pair or in association with a singly charged partner. Search strategies largely depend on the subsequent decays of these scalars, including dileptonic, dibosonic and cascade decay modes. The branching ratios of these decay channels are sensitive to parametric settings within the type-II seesaw models and are mainly determined by the $v_{\Delta}$, the VEV of the triplet scalar and the mass spectra of these scalars. For a relatively large $v_{\Delta}\gg10^{-4}$ GeV, the dibosonic channels $\Delta^{\pm\pm}\to W^{\pm}W^{'\pm}$ dominate; while for a tiny value $v_{\Delta}\ll10^{-4}$ GeV, the dileptonic modes become dominant over the dibosonic ones. For regions in between where the value of $v_{\Delta}$ near $10^{-4}$ GeV, the cascade decays $\Delta^{\pm\pm}\to\Delta^{\pm}W^{\pm*}\to\Delta^{0}W^{\pm*}W^{\pm*}$ are dominant over the other two, if it is allowed kinematically by the mass difference between these triplet components. The dominance of dileptonic decay channels is often assumed in experimental searches, leading to a lower bound on the triplet mass around 1080 GeV by the ATLAS collaboration using 139 fb$^{-1}$ collected events of the Run-2 data at the 13 TeV LHC~\cite{ATLAS:2022pbd}. This result also assumed an equal decay branching ratios of the doubly charged scalar into different lepton flavors. Through dibosonic channels, less stringent limits on $m_{\Delta}$ were given by the ATLAS collaboration around $200\sim220$ GeV with an integrated luminosity of 36 fb$^{-1}$ at the 13 TeV colliding energy~\cite{ATLAS:2018ceg}. A degenerate or nearly degenerate mass spectrum is generally assumed in these searches to forbid the cascade decays, but a mass difference $\Delta m=m_{\Delta^{\pm\pm}}-m_{\Delta^{\pm}}$ is still allowed as large as 40 GeV by the electroweak precision data~\cite{Melfo:2011nx,Chun:2012jw}. Studies in such regions with a non-degenerate mass spectrum bounded the triplet mass from below at $230\sim350$ GeV using the full LHC Run-2 data~\cite{ATLAS:2021jol}. Moreover, the triplet VEV $v_{\Delta}$ itself is subject to restrictions from Electroweak Precision Observables (EWPOs) measurements, as its contributions to radiative corrections on the $\rho$ parameter $\rho\approx1-2v^{2}_{\Delta}/v^{2}$, where $v\approx246$ GeV is the above-mentioned electroweak vacuum expectation value. Global fit results $\rho=1.00031\pm0.00019$~\cite{ParticleDataGroup:2024cfk} thus put an upper bound on the triplet VEV $v_{\Delta}<2.8$ GeV. On the other hand, the lepton-flavor-violating (LFV) rare decays can be mediated via the triplet scalars, such as $\mu^{-}\to e^{-}\gamma$ and $\mu^{-}\to e^{-}e^{-}e^{+}$~\cite{SINDRUM:1987nra,MEG:2016leq}, a limit from below can then be obtained~\cite{Dinh:2012bp,Ashanujjaman:2021txz}
\begin{align}
v_{\Delta}\gtrsim10^{-9}\text{GeV}\times\frac{1\text{TeV}}{m_{\Delta^{\pm\pm}}},
\end{align}
which also depends on the triplet mass $m_{\Delta^{\pm\pm}}$, leading to the rationality of the dileptonic channel dominance assumed in the present work.

% paragraph 3: photon fusion
Besides the traditional Drell-Yan production, the advent of forward detector systems, such as ATLAS Forward Proton Detector~\cite{AFP2015} and CMS-TOTOM~\cite{CTPPS2014}, has unlocked a complementary and powerful investigative channel: elastic photon-photon fusion in ultraperipheral collisions (UPCs) of protons. In these events, the colliding protons interact via their electromagnetic field, approximated as equivalent on-shell photons, and realize initial photon fusion for pair production of charged particles. Protons going through the UPC remain intact and can be detected by the forward facilities, providing a uniquely clean signature characterized by large rapidity gaps between them and the central particles. Hence, even though the photon fusion contribution is shown to be less than that from Drell-Yan production~\cite{Han:2007bk,Fuks:2019clu}, it can offer a powerful tool to suppress overwhelming QCD-dominated backgrounds at hadron colliders and an increasing number of relevant studies have been conducted searching for new physics, such as exotic scalars in seesaw scenarios and left-right symmetric models~\cite{Babu:2016rcr,Duarte:2022xpm,Duarte:2024zeh}, higgsinos in supersymmetric models with compressed spectra~\cite{Godunov:2019jib,Zhou:2022jgj,Zhou:2024fjf} and sleptons as dark matter candidates~\cite{Harland-Lang:2018hmi,Beresford:2018pbt}.

% paragraph 4: this paper
In the present paper, we put forward a promising strategy searching for decays of doubly charged scalars into different-flavor lepton pairs at the high-energy LHC, utilizing elastic photon fusion production in the UPC of protons. The theoretical framework of type-II seesaw accommodating such exotic scalars will be introduced briefly in the next section. In Section 3, we describe the signal and SM background in our simulations. In Section 4 and 5, we present the search strategy and results. Section 6 is our conclusion.

\section{Type-II seesaw model and the triplet scalar decay}

% model content
The generation of neutrino masses via the type-II seesaw mechanism requires an extension of the SM scalar sector. This is achieved by introducing a complex scalar field, extending the SM Lagrangian to include the kinetic and potential terms for the new triplet field~\cite{FileviezPerez:2008jbu,Mandal:2022zmy}
\begin{align}
\mathcal{L}\subset(D_{\mu}H)^{\dagger}(D^{\mu}H)+\text{Tr}[(D_{\mu}\Delta)^{\dagger}(D^{\mu}\Delta)]-V(H,\,\Delta),
\end{align}
in which $H$ is the familiar SM Higgs doublet $H=\left(\phi^{+},\,\phi^{0}\right)^{\text{T}}$ and $\Delta$ the complex scalar transforming as a triplet under the gauge group $SU(2)_{L}$ and lying in the adjoint representation as
\begin{align}
\Delta=
\begin{pmatrix}
\frac{\Delta^{+}}{\sqrt{2}} && \Delta^{++} \\
\Delta^{0} && -\frac{\Delta^{+}}{\sqrt{2}}\,
\end{pmatrix}.
\label{delta}
\end{align}
As the triplet scalar carries a hypercharge of $Y_{\Delta}=1$ within the convention $Q=T_{3}+Y$, the particle content therefore, as presented in Eq.\eqref{delta}, includes doubly charged ($\Delta^{\pm\pm}$), singly charged ($\Delta^{\pm}$) and neutral ($\Delta^{0}$) components. The scalar potential $V(H,\Delta)$ dictates the symmetry-breaking pattern and the masses of the physical scalars, whose general form includes both self-interactions and interactions mixing the doublet and triplet fields~\cite{FileviezPerez:2008jbu,Mandal:2022zmy}
\begin{align}
\label{pot}
V(H,\Delta)={}&-m^{2}_{H}H^{\dagger}H+m^{2}_{\Delta}\text{Tr}[\Delta^{\dagger}\Delta]+[\mu H^{T}i\sigma_{2}\Delta^{\dagger}H+\text{h.c.}]+\frac{\lambda}{4}(H^{\dagger}H)^{2}\notag\\
{}&+\lambda_{1}(H^{\dagger}H)\text{Tr}[\Delta^{\dagger}\Delta]+\lambda_{2}[\text{Tr}(\Delta^{\dagger}\Delta)]^{2}+\lambda_{3}\text{Tr}[(\Delta^{\dagger}\Delta)^{2}]+\lambda_{4}H^{\dagger}\Delta^{\dagger}\Delta H\,,
\end{align}
where $m_{H,\Delta}$ stand for mass parameters of the scalars and $\mu$ is a dimensional coupling for the trilinear term. Dimensionless quartic couplings $\lambda$, $\lambda_{1-4}$ can be taken as real numbers without loss of generality. After the spontaneous electroweak symmetry breaking, the neutral components of both the doublet and the triplet acquire vacuum expectation values (VEVs): $\langle H\rangle=v_{0}$ and $\langle\Delta\rangle=v_{\Delta}$. Minimization of the potential Eq.\eqref{pot} leads to the relation $v_{\Delta}\approx\mu v^{2}_{0}/m^{2}_{\Delta}$, revealing how a tiny triplet VEV $v_{\Delta}$ naturally emerges from a large scalar mass $m_{\Delta}$ and$\backslash$or a small coupling $\mu$ while the electroweak vacuum remains $v^{2}=v^{2}_{0}+v^{2}_{\Delta}\approx246^{2}\,\text{GeV}^{2}$.

% decay and neutrino mass matrix
The connection to neutrino physics arises from a new Yukawa interaction introduced in the type-II seesaw model. This interaction couples the triplet scalar directly to the SM leptons
\begin{align}
\mathcal{L}_{Y}\subset-y_{\Delta}L^{T}Ci\sigma_{2}\Delta L+\text{h.c.}\,,
\label{yukawa}
\end{align}
in which $L=(\nu_{\ell}\,,\ell)^{T}$ is the left-handed lepton doublets, $y_{\Delta}$ is the Yukawa coupling matrix in flavor space and $C$ the charge conjugation operator. Upon electroweak symmetry breaking, this interaction generates in a seesaw style Majorana masses for the neutrinos $\mathcal{M}_{\nu}=\sqrt{2}Y_{\Delta}v_{\Delta}$, which can be naturally tiny for the same reason as explained above for the triplet scalar VEV $v_{\Delta}$. The mass matrix $\mathcal{M}_{\nu}$ is most generally diagonalized by the Pontecorvo-Maki-Nakagawa-Sakata (PMNS) matrix~\cite{ParticleDataGroup:2024cfk}, such that $\mathcal{M}_{\nu}=U^{*}\text{diag}(m_{1},m_{2},m_{3})U^{\dag}$, establishing a profound and testable link between low-energy neutrino oscillation parameters and high-energy collider signatures. Current neutrino oscillation data can fix to a great extent the elements of the PMNS matrix, including the three mixing angles $\theta_{12},\theta_{13},\theta_{23}$ and the Dirac CP phase $\delta$~\cite{Esteban:2024eli,nufit}. However, as the oscillation experiments are only sensitive to the mass difference squared and Dirac phase, two Majorana phases $\eta_{1,2}$ and the minimal neutrino mass remain undetermined. This also leads to two kinds of neutrino mass hierarchies, commonly known as the normal or inverted ones, corresponding to $\nu_{1}$ or $\nu_{3}$ as the lightest neutrino. Due to the Yukawa interaction in~\eqref{yukawa}, the doubly charged triplet scalars $\Delta{^{\pm\pm}}$ are possible to decay into a pair of same-sign leptons, the branching ratios of which are dictated by the neutrino mass matrix $\mathcal{M}_{\nu}$ and the mass hierarchy - normal (NH) or inverted (IH), offering a unique fingerprint of the underlying neutrino physics in collider events. As mentioned in Section~\ref{sec:intro}, the overall dominance of the decay modes is highly sensitive the VEV of the triplet $v_{\Delta}$, which can be seen clearly from its dileptonic decay width~\cite{Melfo:2011nx,Chun:2003ej,FileviezPerez:2008jbu,Mandal:2022zmy}
\begin{align}
\label{lepbr}
\Gamma(\Delta^{\pm\pm}\to\ell^{\pm}_{\alpha}\ell^{\pm}_{\beta})=\frac{m_{\Delta^{\pm\pm}}}{8\pi(1+\delta_{\alpha\beta})}\left|\frac{\mathcal{M}^{ij}_{\nu}}{v_{\Delta}}\right|^{2},
\end{align}
with $\alpha$, $\beta$ being the lepton flavors and $\delta_{\alpha\beta}$ the Kronecker symbol. An extremely small value of $v_{\Delta}<10^{-4}$ GeV can significantly increase the ratios of the dileptonic channels and, at the same time, highly suppress the decay ratios into dibosons
\begin{align}
\Gamma(\Delta^{\pm\pm}\to W^{\pm}W^{\pm})=\frac{g^{4}v^{2}_{\Delta}}{8\pi m_{\Delta^{\pm\pm}}}\sqrt{1-\left(\frac{2m_{W}}{m_{\Delta^{\pm\pm}}}\right)^{2}}\left[\left(\frac{m^{2}_{\Delta^{\pm\pm}}}{2m^{2}_{W}}-1\right)^{2}+2\right]\,,
\end{align}
where $g$ is the weak gauge coupling.

For this study, we focus on the regime of a small $v_{\Delta}$ and a degenerate scalar mass spectrum, which is well-motivated by constraints from electroweak precision tests (which favor a small $v_{\Delta}$ as discussed in Section~\ref{sec:intro}) and ensures the dominance of the clean, fully leptonic decay channels most accessible at the LHC. Using the best-fit values for neutrino oscillation parameters - including mixing angles, mass-squared differences, and the CP-violating phase - the Yukawa coupling matrix is constrained. This, in turn, determines the branching fractions of $\Delta^{\pm\pm}$ into specific lepton pairs. These best-fit values are adopted from the NuFIT program~\cite{Esteban:2024eli,nufit} for both the normal hierarchy (NH) and inverted hierarchy (IH):
\begin{align}
\label{pmns:NH}
\text{NH}: \quad{}&\Delta m^{2}_{21}=7.49\times10^{-5}\text{eV}^{2},\,\,\Delta m^{2}_{31}=2.513\times10^{-3}\text{eV}^{2},\notag\\ {}&\sin^{2}\theta_{12}=0.308,\,\,\sin^{2}\theta_{23}=0.470,\,\,\sin^{2}\theta_{13}=0.02215,\,\, \delta_{CP}=212^{\circ},\\
\label{pmns:IH}
\text{IH}:\quad{}&\Delta m^{2}_{21}=7.49\times10^{-5}\text{eV}^{2},\,\,\Delta m^{2}_{31}=-2.484\times10^{-3}\text{eV}^{2},\notag\\ {}&\sin^{2}\theta_{12}=0.308,\,\,\sin^{2}\theta_{23}=0.550,\,\,\sin^{2}\theta_{13}=0.02231,\,\, \delta_{CP}=274^{\circ}.
\end{align}
In addition, the lightest neutrino mass is taken as $0.05$ eV and vanishing Majorana phases $\eta_{1,2}=0$ are assumed in the present work. Under these assumptions and data input, we will focus on the LFV decay channel of $\Delta^{\pm\pm}\to e^{\pm}\mu^{\pm}$ whose branching ratio approximates to a level of 1\% or even smaller, which will be discussed in detail in the following sections. As a complementary study to the same-flavor dileptonic channels $\Delta^{\pm\pm}\to e^{\pm}e^{\pm}/\mu^{\pm}\mu^{\pm}$ in~\cite{Zhou:2025ljo}, we will demonstrate in the following sections that although the different-flavor branching ratio is far less than that of same-flavor modes (Br$(\Delta^{\pm\pm}\to\mu\mu)\approx25\%$ for the case of NH and Br$(\Delta^{\pm\pm}\to ee)\approx47\%$ for the case of IH), a promising probing sensitivity can also be achieved at the high-energy LHC.

\section{Production and signatures at the LHC}
\label{sec:signature}

\begin{figure}[t]
\centering
\includegraphics{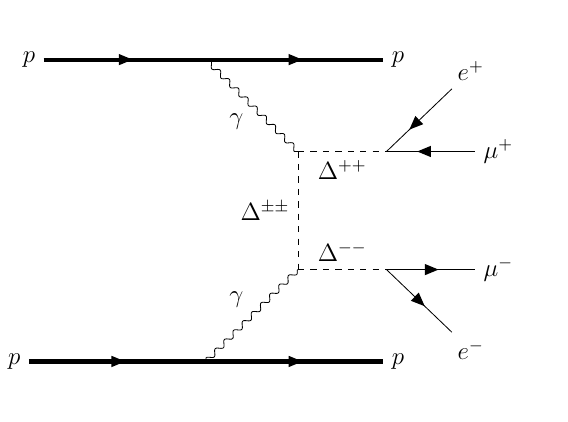}
\caption{Feynman diagrams for the signal process of lepton-flavor-violating decay channel from doubly charged scalars pair production through elastic photon fusion at the LHC: $pp\to p(\gamma\gamma\to\Delta^{++}\Delta^{--}\to e^{+}\mu^{+}e^{-}\mu^{-})p$\,.}
\label{fig:sigFD}
\end{figure}

As mentioned in Section~\ref{sec:intro}, the search strategy employed in this analysis pivots on a distinctive production mechanism: the pair production of doubly charged scalars via the fusion of elastic photons in ultraperipheral proton collisions. This process, conceptually illustrated in the Feynman diagram in Figure.~\ref{fig:sigFD}:
\begin{align}
pp\to p(\gamma\gamma\to\Delta^{++}\Delta^{--})p\,,
\end{align}
offers a signature that is remarkably clean compared to standard Drell-Yan production channels. Feynman diagrams in the present paper are drawn via the TikZ-Feynman package~\cite{Ellis:2016jkw}. The defining characteristic is the presence of the two colliding protons in the final state, which remain intact after radiating the interacting photons elastically and are detectable by dedicated forward detectors. This exclusive process can be treated within the equivalent photon approximation (EPA), the framework of which approximates the electromagnetic fields of the high-energy protons as fluxes of on-shell photons, characterized by photon parton distribution functions $\gamma-$PDF. The total cross section can then be expressed in the form:
\begin{align}
\sigma_{pp\to p(\gamma\gamma\to\Delta^{++}\Delta^{--})p}=\int\sigma_{\gamma\gamma\to\Delta^{++}\Delta^{--}}f_{\gamma/p}(z_{1})f_{\gamma/p}(z_{2})dz_{1}dz_{2}\,,
\end{align}
in which $f_{\gamma/p}(z_{1,2})$ describes the probability for a proton to radiate an elastic photon carrying a momentum fraction $z_{1,2}$, and $\sigma_{\gamma\gamma\to\Delta^{++}\Delta^{--}}$ is the hard subprocess cross-section. Crucially, the integration limits are constrained by the acceptance of the forward detectors, which is a function of the proton's fractional energy loss $\xi\equiv1-E_{\text{out}}/E_{\text{in}}$. That is, the energy of outgoing protons ($E_{\text{out}}$) and that of incident protons ($E_{\text{in}}$) determine the energy loss and then the detection efficiencies of the intact protons by the forward facilities. The energy loss, considering the elastic nature of the UPC process, can be taken as equal to the energy of the EPA photon $E_{\gamma}$ emitted from the incoming protons. Based on documented detector performance, a set of efficiency values was adopted for different photon energy intervals: for $E_{\gamma}$ from 100 GeV to 1 TeV, the detection rate for final protons approximates 100\% at the 13 TeV LHC, which can be translated to the energy loss $\xi$ of (0.015, 0.15)~\cite{CTPPS2014,AFP2015}. In our following simulations in the next section, we adopt more conservative detection rates (Table.~\ref{tab:protonrates}) since lower efficiencies around 90\% are generally indicated by phenomenological studies~\cite{Beresford:2018pbt}.

\begin{table}
%\flushleft
\centering
\begin{tabular}{|c|*{5}{c|}}
\hline
\,$E_{\gamma}$\,(GeV)\, & \,(0,100]\, & \,(100,120]\, & \,(120,150]\, & \,(150,400]\, & \,(400,$+\infty$)\, \\ \hline
Eff. & 0 & 50\% & 70\% & 90\% & 80\% \\ \hline
\hline
\end{tabular}
\caption{Acceptance rates for initial photons with different ranges of energies, which are equivalent to tagging efficiencies for the outgoing protons corresponding to their energy losses \cite{CTPPS2014,AFP2015}.}
\label{tab:protonrates}
\end{table}

As discussed in the last section, we decay the doubly charged scalars subsequently into lepton pairs of different flavors: $\Delta^{\pm\pm}\to e^{\pm}\mu^{\pm}$, with the assumption of $v_{\Delta}<10^{-4}$ GeV and hence the dileptonic channel dominance. The signal process can then be written as
\begin{align}
pp\to p(\gamma\gamma\to\Delta^{++}\Delta^{--}\to e^{+}\mu^{+}e^{-}\mu^{-})p\,,
\label{proc:sig}
\end{align}
the final state of which includes two intact protons and two pairs of same-sign leptons $e^{\pm}\mu^{\pm}$ coming from a pair of the exotic scalars. In the normal and inverted mass hierarchies (Eq.~\eqref{pmns:NH} and \eqref{pmns:IH}), the decay $\Delta^{\pm\pm}\to e^{\pm}\mu^{\pm}$ has a low branching ratio in both cases: Br$(\Delta^{\pm\pm}\to e^{\pm}\mu^{\pm})\approx$0.5\% for NH and 2.6\% for IH.

\begin{figure}[t]
\centering
\begin{minipage}{0.38\linewidth}
  \centerline{\includegraphics[scale=0.78]{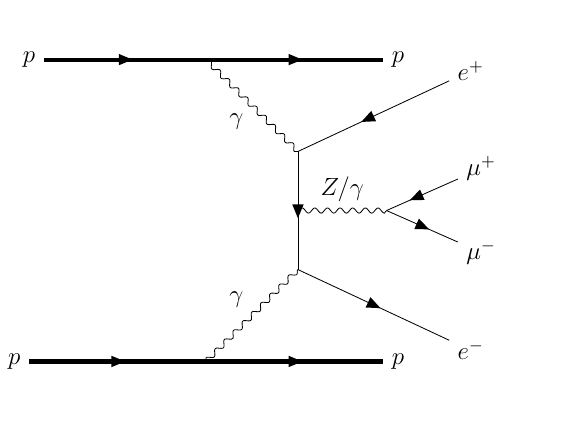}}
  \centerline{(a)}
\end{minipage}
\qquad\qquad
\begin{minipage}{0.38\linewidth}
  \centerline{\includegraphics[scale=0.78]{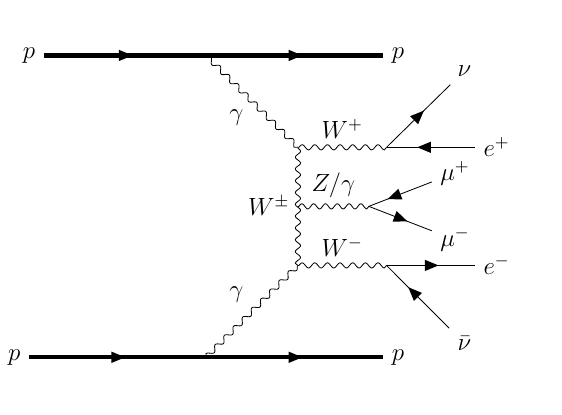}}
  \centerline{(b)}
\end{minipage}
\\[12pt]
\begin{minipage}{0.38\linewidth}
  \centerline{\includegraphics[scale=0.78]{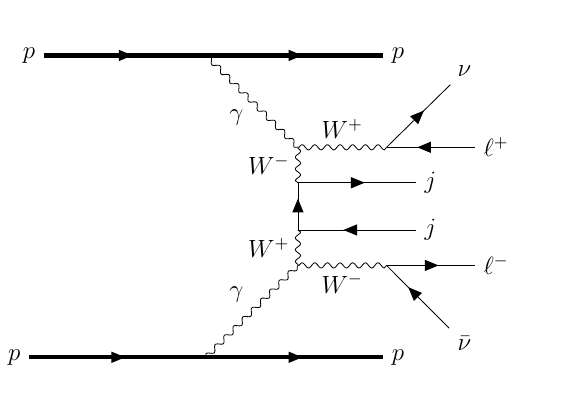}}
  \centerline{(c)}
\end{minipage}
\qquad\qquad
\begin{minipage}{0.38\linewidth}
  \centerline{\includegraphics[scale=0.78]{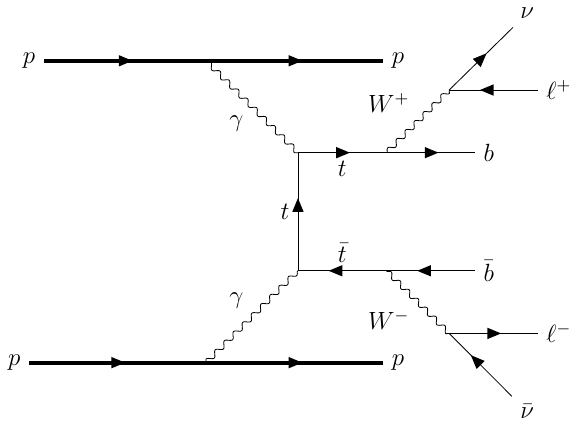}}
  \centerline{(d)}
\end{minipage}
\caption{Feynman diagrams of the Standard Model background for fully leptonic channels through elastic photon fusion at the LHC. $t$-channel $Z/\gamma$-mediating diagrams in (a) and (b) are not shown but taken into consideration in our analysis. Final lepton flavors $e$ and $\mu$ in (a) and (b) can be flipped, while the lepton flavors $\ell$ in (c) and (d) are not shown explicitly for simplicity, as they can form several flavor combinations with the misidentified jets.}
\label{fig:bkgFD}
\end{figure}

The corresponding Standard Model backgrounds that mimic this signature are processes that can yield four prompt, isolated leptons in conjunction with two intact forward protons via photon fusion. Direct production of a pair of opposite-sign leptons from photon fusion contributes to irreducible backgrounds if an associated $Z$ boson or photon is produced, decaying into a lepton pair
\begin{align}
\label{proc:bkg1}
pp\to{}&p(\gamma\gamma\to\ell^{+}_{\alpha}\ell^{-}_{\alpha}Z/\gamma\to\ell^{+}_{\alpha}\ell^{-}_{\alpha}\ell^{+}_{\beta}\ell^{-}_{\beta})p\,,
\end{align}
where lepton pairs from photon fusion $\ell^{+}_{\alpha}\ell^{-}_{\alpha}$ and from associated boson $\ell^{+}_{\beta}\ell^{-}_{\beta}$ are of different flavors $\alpha=e$, $\beta=\mu$ or vice versa, as shown in \figurename.~\ref{fig:bkgFD}(a). Another background comes from the pair production of $W^{+}W^{-}$ from photon fusion associated with a $Z/\gamma$, also decaying into a lepton pair
\begin{align}
\label{proc:bkg2}
pp\to{}&p(\gamma\gamma\to W^{+}W^{-}Z/\gamma\to\ell^{+}_{\alpha}\nu\ell^{-}_{\alpha}\bar{\nu}\ell^{+}_{\beta}\ell^{-}_{\beta})p\,.
\end{align}
where leptons from $W$ decays and from associated $Z/\gamma$ are of different flavors as in the above case of Eq.~\eqref{proc:bkg1}, corresponding to \figurename.~\ref{fig:bkgFD}(b). Alongside the pair production of $W$ bosons, if two light jets are produced associatively from the mediating $W$ boson and misidentified as leptons, this process also contributes as an irreducible background
\begin{align}
\label{proc:bkg3}
pp\to&p(\gamma\gamma\to W^{+}W^{-}jj\to\ell^{+}_{\alpha}\nu\ell^{-}_{\alpha}\bar{\nu}jj)p\,,
\end{align}
in which the leptons from $W$ pairs are of the same flavor $\alpha$ and two jets are misidentified as another flavor $\beta$ which is not shown explicitly (\figurename.~\ref{fig:bkgFD}(c)). Finally, we consider top pair production from photon fusion with two top quarks decaying into $b$-jets, leptons and neutrinos
\begin{align}
\label{proc:bkg4}
pp\to&p(\gamma\gamma\to t\bar{t}\to b\ell^{+}_{\alpha}\nu\bar{b}\ell^{-}_{\alpha}\bar{\nu})p\,,
\end{align}
where, similar as the third background Eq.~\eqref{proc:bkg3}, two $b$-jets are misidentified as same-flavor opposite-sign leptons $\ell_{\beta}$ (\figurename.~\ref{fig:bkgFD}(d)).

\begin{comment}
\begin{table}
%\flushleft
\centering
\begin{tabular}{|c|*{4}{c|}}
\hline
\diagbox{$\sqrt{s}$}{BKG} & $\ell^{+}\ell^{-}Z/\gamma$ & $W^{+}W^{-}Z/\gamma$ & $W^{+}W^{-}jj$ & $t\bar{t}$ \\ \hline
14 TeV & $8.523\times10^{-6}$ & $4.684\times10^{-7}$ & $2.552\times10^{-3}$ & $7.705\times10^{-6}$ \\ \hline
100 TeV & $3.258\times10^{-5}$ & $1.194\times10^{-5}$ & $2.289\times10^{-2}$ & $8.708\times10^{-5}$ \\ \hline
\hline
\end{tabular}
\caption{Cross sections of SM background at the 14 and 100 TeV LHC, corresponding to the processes from Eq.~\eqref{proc:bkg1} to Eq.~\eqref{proc:bkg4}. Cross sections are in unit of picobarn.}
\label{tab:bkgxs}
\end{table}
\end{comment}

\begin{table}
%\flushleft
\centering
\begin{tabular}{|c|*{4}{c|}}
\hline
\diagbox{$\sqrt{s}$}{BKG} & $\ell^{+}\ell^{-}Z/\gamma$ & $W^{+}W^{-}Z/\gamma$ & $W^{+}W^{-}jj$ & $t\bar{t}$ \\ \hline
$\sigma$ at 100 TeV & $3.258\times10^{-5}$ & $1.194\times10^{-5}$ & $2.289\times10^{-2}$ & $8.708\times10^{-5}$ \\ \hline
\hline
\end{tabular}
\caption{Cross sections of SM background at the 100 TeV LHC, corresponding to the processes from Eq.~\eqref{proc:bkg1} to Eq.~\eqref{proc:bkg4}. Cross sections are in unit of picobarn.}
\label{tab:bkgxs}
\end{table}

\begin{figure}[t]
\centering
\includegraphics[scale=0.95]{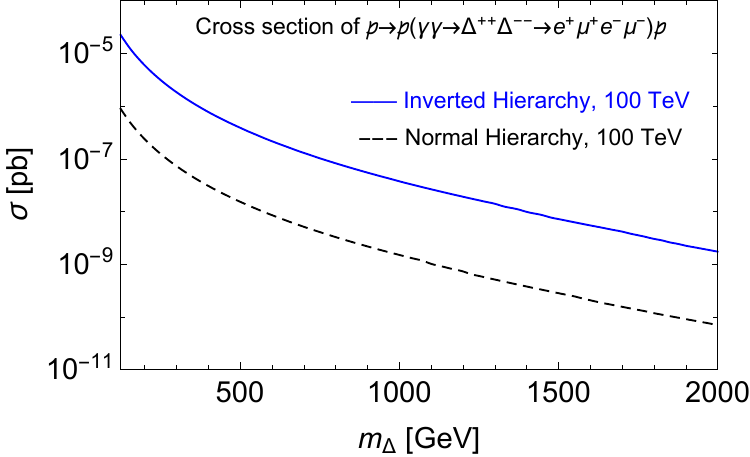}
\caption{Cross sections of the signal process versus the triplet scalar mass at the 100 TeV LHC. Normal and inverted hierarchies of the neutrino mass are both displayed in the figure.}
\label{fig:sigxs}
\end{figure}

We consider the above signal and background processes in a high colliding energies of 100 TeV at the LHC. In Table.~\ref{tab:bkgxs} we present the cross sections of the four background processes corresponding to Eq.~\eqref{proc:bkg1} $\sim$ Eq.~\eqref{proc:bkg4}. Cross sections of the signal processes are shown in Figure.~\ref{fig:sigxs} varying from $10^{-11}$ to $10^{-5}$ pb with regarding to the triplet scalar mass ranging from 125 to 2000 GeV. The distinction between the cross sections in NH and IH cases can be attributed to the tiny difference between the LFV decay branching ratios of $\Delta^{\pm\pm}\to e^{\pm}\mu^{\pm}$ under these two cases as mentioned above.

\section{Simulation and search strategies}
\label{sec:search}

To quantitatively assess the discovery potential of the proposed search channel, we have conducted a comprehensive Monte Carlo simulation study encompassing both signal and background processes. The event generation pipeline begins at the parton level using the \textsc{MadGraph5\_aMC@NLO} of version 3.5.6~\cite{Alwall2014}, which provides leading-order matrix elements for all relevant processes. For the signal process, we employ the \textsc{Typeiiseesaw} Universal \textsc{Feynrules} Output (UFO) libraries~\cite{Fuks:2019clu}, which encode the complete Lagrangian of the type-II seesaw scenario, including all interactions necessary for accurate simulation of doubly charged scalar production and decay. For the photon-initiated nature of the production mechanism, we utilize the
$\gamma$-UPC packages~\cite{Shao:2022cly} developed for describing equivalent photon fluxes in ultraperipheral collisions, which implements the elastic photon parton distribution functions for protons based on established analytic parameterizations. Following parton-level generation, events are passed through \textsc{Pythia-8.2}~\cite{Sjostrand:2014zea} for parton showering and hadronization. Detector effects are simulated using the \textsc{Delphes-3.5.0}~\cite{deFavereau:2013fsa} fast simulation framework, with detector parameters configured to match the performance of the ATLAS detectors at the LHC. The entire simulation chain, including the parton-level event generation, parton showering and detector simulation, is managed within the \textsc{CheckMATE2} framework~\cite{Dercks2017}.

\begin{figure}
\centering
\begin{minipage}{0.38\linewidth}
  \centerline{\includegraphics[scale=0.45]{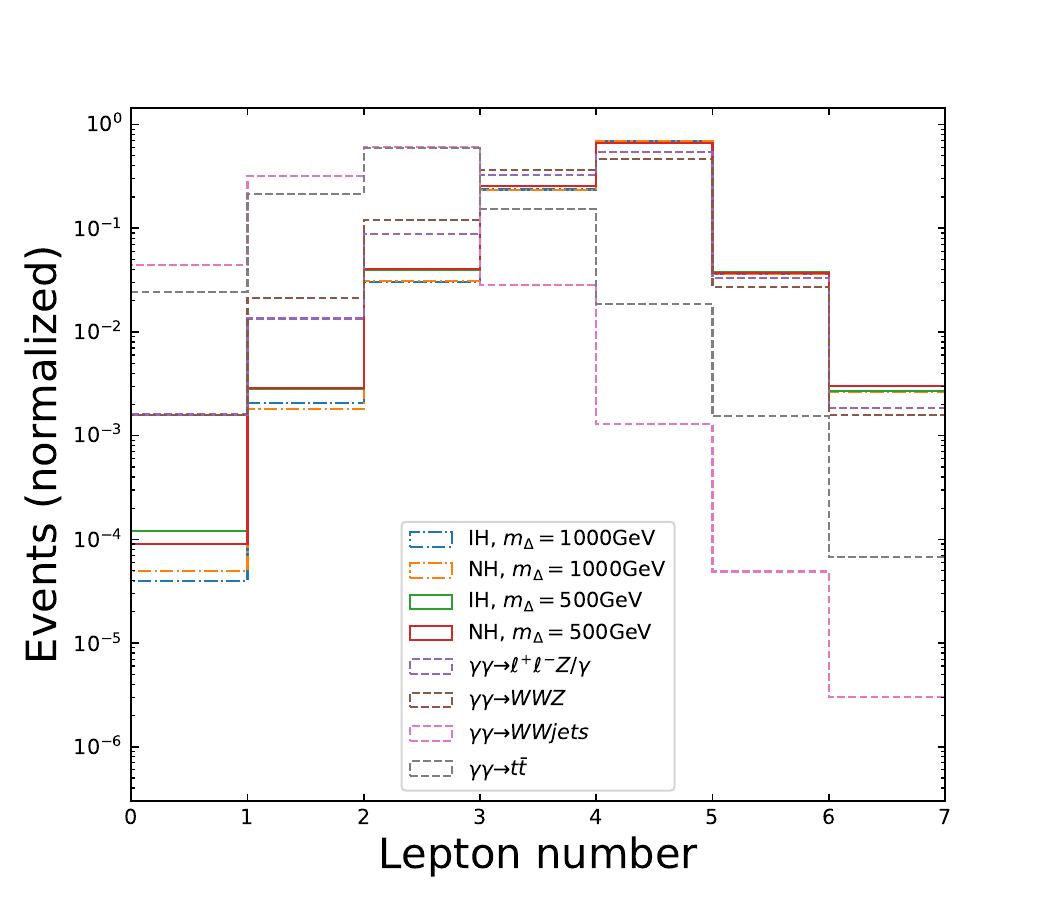}}
  \centerline{(a)}
\end{minipage}
\qquad\qquad
\begin{minipage}{0.38\linewidth}
  \centerline{\includegraphics[scale=0.45]{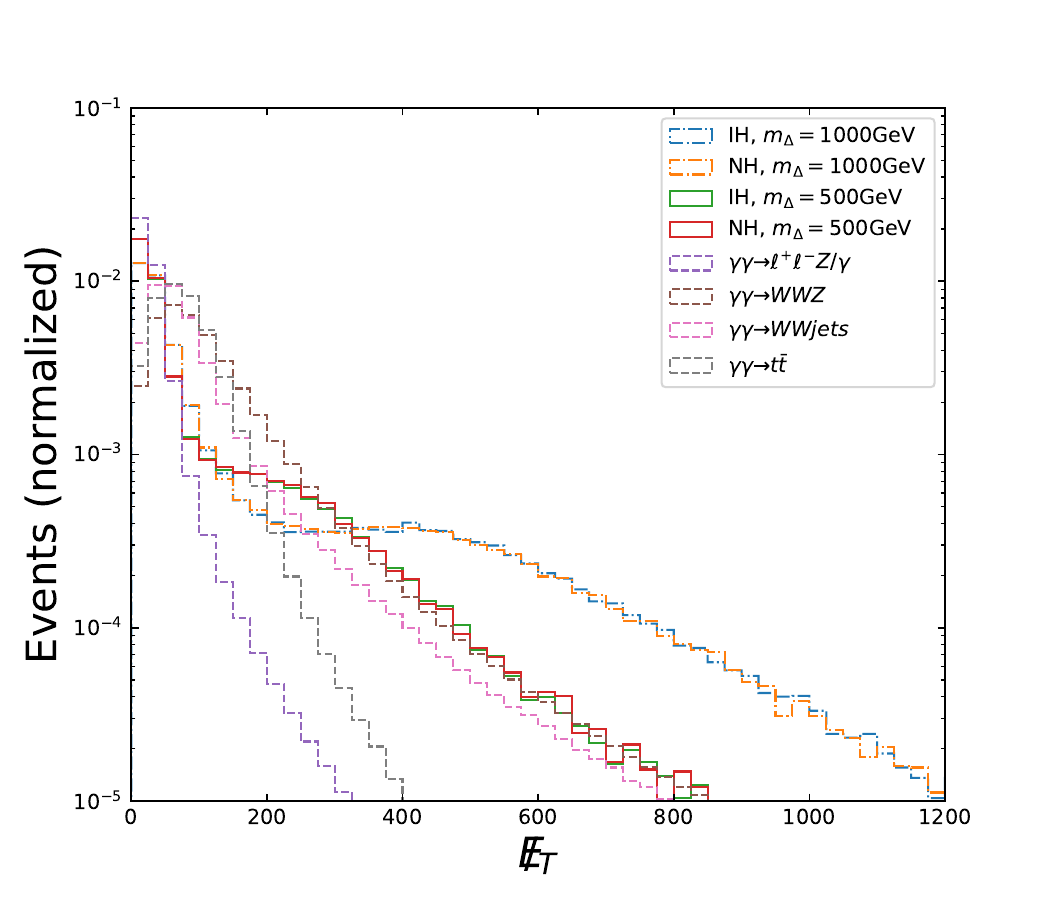}}
  \centerline{(b)}
\end{minipage}
%\\[1pt]
\begin{minipage}{0.38\linewidth}
  \centerline{\includegraphics[scale=0.45]{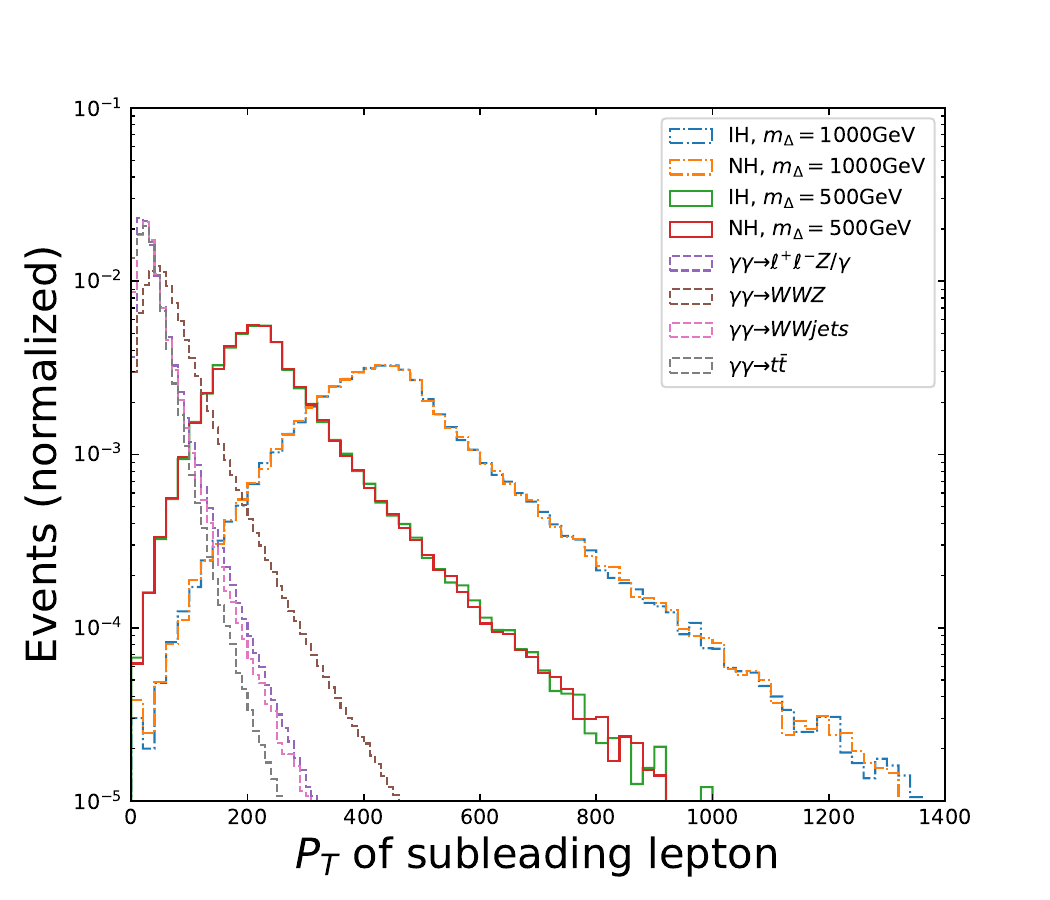}}
  \centerline{(c)}
\end{minipage}
\qquad\qquad
\begin{minipage}{0.38\linewidth}
  \centerline{\includegraphics[scale=0.45]{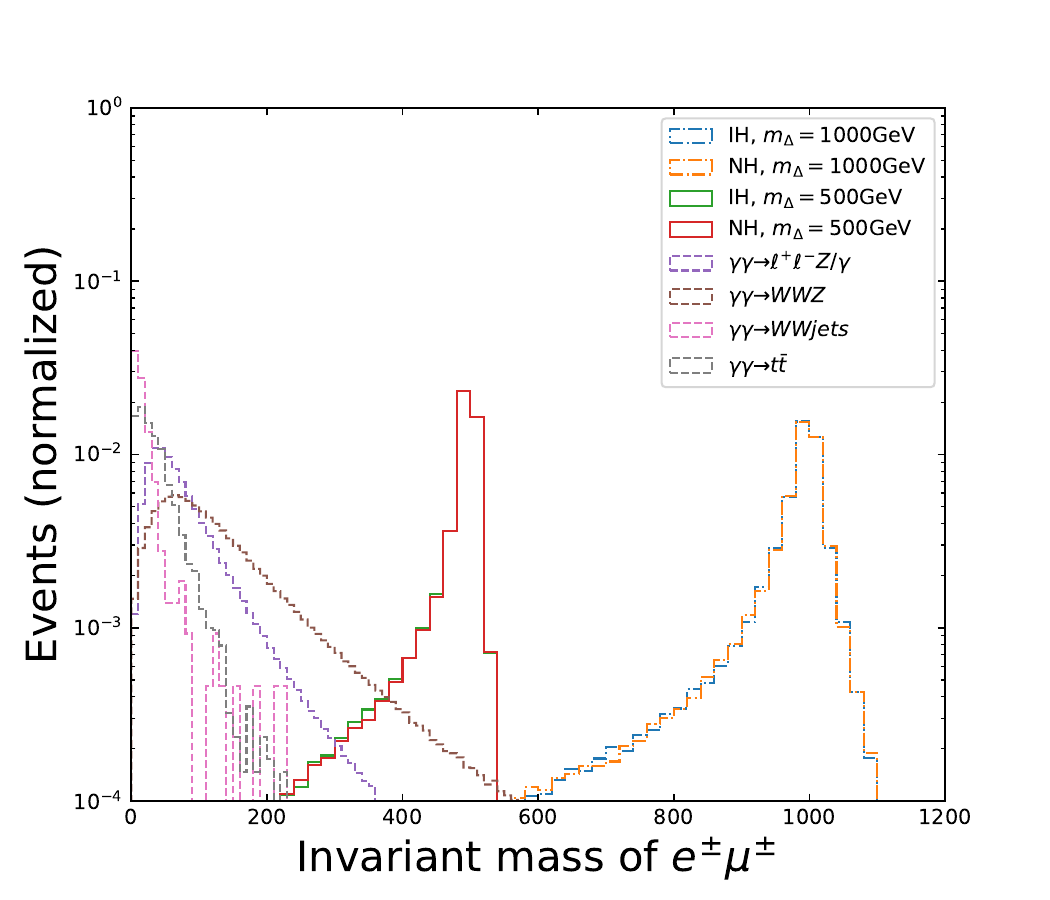}}
  \centerline{(d)}
\end{minipage}
\caption{Normalized distributions for kinematical variables including lepton number $N_{\ell}$, missing transverse energy $\slashed{E}_{T}$, transverse momentum of the subleading lepton $P_{T}(\ell_{2})$ and invariant mass of same-sign different-flavor lepton pair $m(e^{\pm}\mu^{\pm})$ from elastic photon fusion at the 100 TeV LHC: $pp\to p(\gamma\gamma\to\Delta^{++}\Delta^{--}\to e^{+}\mu^{+}e^{-}\mu^{-})p$\,. Two benchmarks of $m_{\Delta}=500$ and 1000 GeV under both normal and inverted hierarchies are displayed, with solid (500 GeV) and dash-dotted lines (1000 GeV). Distributions for four SM background processes are shown in dashed lines.}
\label{fig:dist}
\end{figure}

The generated events exhibit distinct kinematic features that enable further discrimination between signal and background. Figure.~\ref{fig:dist} illustrates the normalized distributions of four powerful discriminants, comparing signal events under benchmarks of $m_{\Delta}=500$ (solid lines) and 1000 GeV (dash-dotted lines) against the four SM backgrounds (dashed lines), including lepton number $N_{\ell}$, missing transverse energy $\slashed{E}_{T}$, transverse momentum of the subleading lepton $P_{T}(\ell_{2})$, invariant mass reconstructed from the same-sign different-flavor lepton pair $e^{\pm}\mu^{\pm}$. Normal and inverted hierarchies for neutrino mass spectra are also displayed for each of the two benchmarks. Lepton number distributions for signal processes peak at $N_{\ell}=4$, while the $\gamma\gamma\to\ell^{+}\ell^{-}Z/\gamma$ and $\gamma\gamma\to WWZ$ events have similar distributions centering around $N_{\ell}=4$ as expected (Figure.~\ref{fig:dist}(a)). Missing transverse energy can reveal further distinguishable features between signal and background, especially for benchmarks of a larger triplet mass $m_{\Delta}$. As shown in \figurename.~\ref{fig:dist}(b), both signal and background events tend to distribute in a range of smaller $\slashed{E}_{T}$, but the signal curves decline more sharply as $\slashed{E}_{T}$ increases and this feature becomes clearer for larger $m_{\Delta}$. Even more powerful discriminants are lepton $P_{T}$ and the reconstructed invariant mass. Leptons from a larger-mass particle decay tend to possess larger momenta, while the momenta of leptons in background events are relatively smaller. For simplicity, we only present the subleading lepton $P_{T}$ in \figurename.\ref{fig:dist}(c), the kinematic distributions for the leading lepton are similar. In \figurename.\ref{fig:dist}(d), we can see clear peaks for the reconstructed mass from $e^{\pm}\mu^{\pm}$ pairs in signal events around the parent particle mass 500 and 1000 GeV, which can be used as an efficient cut to suppress the background. It can also be seen from the histograms in \figurename.\ref{fig:dist} that, for two cases of normal and inverted hierarchies, the events exhibit very similar distributions of the considered kinematic variables, which is a natural result since these kinematic distributions are much more sensitive to the mass of the decay parent particle $\Delta^{\pm\pm}$, than to the absolute values of the decay branching ratios determined in part by the neutrino mass hierarchy. One can then expect similar probing sensitivities for these two hierarchies, which we will discuss in the next section.

\begin{table}{b}
%\flushleft
\centering
\begin{tabular}{|c|*{4}{c|}}
\hline
\diagbox{Cut-4}{SR} & $m_{\Delta}<600$ & $m_{\Delta}\in(600,900]$ & $m_{\Delta}\in(900,1250]$ & $m_{\Delta}>1250$\qquad \\ \hline
$m(e^{\pm}\mu^{\pm})\in$ & $m_{\Delta}\pm10$ & $m_{\Delta}\pm20$ & $m_{\Delta}\pm52$ & $m_{\Delta}\pm110$\\ \hline
\hline
\end{tabular}
\caption{Cut-4 in the event selections based on signal regions (SR) related to the triplet mass $m_{\Delta}$. For simplicity, we use $m_{\Delta}\pm10$ in short for the range of $[m_{\Delta}-10, m_{\Delta}+10]$, so that for SR of $m_{\Delta}<600$ GeV for example, the cut-4 is to require the invariant mass $m(e^{\pm}\mu^{\pm})\in[m_{\Delta}-10, m_{\Delta}+10]$, etc. Values of mass are in unit of GeV.}
\label{tab:cut-4}
\end{table}

All the above kinematic features motivate the following event selection criteria to realize the optimal sensitivity:
\begin{itemize}
\item Pre-selection: Two intact outgoing protons.
\item Cut-1: Two pairs of same-sign different-flavor leptons: $e^{+}\mu^{+}e^{-}\mu^{-}$.
\item Cut-2: Transverse missing energy $\slashed{E}_{T}<180$ GeV.
\item Cut-3: Transverse momenta of the subleading leptons $P_{T}(\ell_{2})>m_{\Delta}/8$.
\item Cut-4: Invariant mass of same-sign different-flavor lepton pairs $m(e^{\pm}\mu^{\pm})$ based on signal regions related to $m_{\Delta}$ in \tablename.~\ref{tab:cut-4}.
\end{itemize}
where we also list the requirement of tagging two intact outgoing protons as the very first cut, acting as a pre-selection procedure, since the protons go through fully elastic scattering and are to be tagged by the forward detectors with energy-loss-dependent efficiencies discussed above in Section.~\ref{sec:signature} and \tablename.~\ref{tab:protonrates}. Two pairs of same-sign different-flavor leptons and a relatively small missing transverse energy are then required as cut-1 and cut-2. At least two of these leptons are required to have large enough transverse momenta depending on their parent particle mass, for which we apply cut-3 for the subleading lepton transverse momentum $p_{T}(\ell_{2})>m_{\Delta}/8$. Finally, the invariant mass reconstructed from the same-sign different-flavor lepton pair $m(e^{\pm}\mu^{\pm})$ should lie in a range around the triplet scalar mass. For better probing sensitivity in the mass range from 125 to 2000 GeV, we consider four signal regions (SR) for the cut on $m(e^{\pm}\mu^{\pm})$. In the range of a few hundreds GeV, the reconstructed invariant mass for $e^{\pm}\mu^{\pm}$ pair is not well separated from the SM backgrounds in which the leptons come from $W$ or $Z$ bosons, which can also be seen from the distributions for $m(e^{\pm}\mu^{\pm})$ in \figurename.~\ref{fig:dist}(d). As the triplet mass becomes larger, the peak of $m(e^{\pm}\mu^{\pm})$ moves further from the background peaks. In consideration of this behavior, we adopt a narrow mass interval for $m(e^{\pm}\mu^{\pm})$ cut between $m_{\Delta}\pm10$ GeV in the SR of $m_{\Delta}<600$ GeV, and the interval is increased to 20, 52 and 110 GeV for SRs of larger triplet mass, see \tablename.~\ref{tab:cut-4} for details.

\begin{table}
\centering
\begin{tabular}{|l|c|c|c|c|c|}
\hline
  & \makecell{$m_{\Delta}=1150$ \\ IH} &  $\ell^{+}\ell^{-}Z/\gamma$ & $W^{+}W^{-}Z/\gamma$ & $W^{+}W^{-}jj$ & $t\bar{t}$ \\ \hline
No cuts & $2.22\times10^{-8}$ & $3.26\times10^{-5}$ & $1.19\times10^{-5}$ & $2.29\times10^{-2}$ & $8.71\times10^{-5}$ \\\hline
2 protons & $1.43\times10^{-8}$  & $5.04\times10^{-6}$ & $5.92\times10^{-6}$ & $4.97\times10^{-3}$ & $2.54\times10^{-5}$ \\ \hline
$e^{+}\mu^{+}e^{-}\mu^{-}$ & $8.96\times10^{-9}$ & $1.64\times10^{-6}$ & $5.15\times10^{-7}$ & $1.65\times10^{-6}$ & $9.68\times10^{-8}$ \\ \hline
Low $\slashed{E}_{T}$ & $6.76\times10^{-9}$ & $1.57\times10^{-6}$ & $2.71\times10^{-7}$ & $1.10\times10^{-6}$ & $8.92\times10^{-8}$ \\ \hline
High $P_{T}(\ell_{2})$ & $6.03\times10^{-9}$ & $1.24\times10^{-8}$ & $2.11\times10^{-9}$ & $0$& $0$ \\ \hline
$m(e^{\pm}\mu^{\pm})$ & $3.19\times10^{-9}$ & $6.52\times10^{-12}$ & $0$ & $0$& $0$ \\
\hline
\end{tabular}
\caption{Cutflow for effective cross sections of the signal process $pp\to p(\gamma\gamma\to\Delta^{++}\Delta^{--}\to e^{+}\mu^{+}e^{-}\mu^{-})p$ under IH neutrino mass spectrum with $m_{\Delta}=1150$ GeV, and of four SM backgrounds $\ell^{+}\ell^{-}Z/\gamma$, $W^{+}W^{-}Z/\gamma$, $W^{+}W^{-}jj$ and $t\bar{t}$ events at the 100 TeV LHC. Cross sections and masses are in units of picobarn and GeV, respectively, which are omitted in the table for simplicity.}
\label{tab:cutflow}
\end{table}

To illustrate the effectiveness of this sequential selection strategy, \tablename.~\ref{tab:cutflow} presents detailed cutflows for a representative benchmark point $m_{\Delta}=1150$ GeV and the SM backgrounds. Only the case of IH is presented in the table for cutflow, since the resulting sensitivity for the NH case is less promising than that for the IH case, which we will discuss in the next section. The forward proton tagging requirement reduces all exclusive processes by approximately a half. The subsequent requirement of exactly four leptons with the appropriate charge and flavor combinations proves remarkably effective against the $W^{+}W^{-}+$jets and $t\bar{t}$ backgrounds, which are almost completely eliminated due to their inherent lack of four isolated leptons. The $P_{T}$ cut on the subleading lepton provides moderate additional suppression while retaining most signal events. The most dramatic background reduction comes from the invariant mass cut $m(e^{\pm}\mu^{\pm})$. This selection retains approximately 50\% of signal events while reducing the $\ell^{+}\ell^{-}Z/\gamma$ and $W^{+}W^{-}Z/\gamma$ backgrounds to zero or to a negligible level. The other two backgrounds are completely eliminated, as they lack any mechanism to produce a same-sign lepton pair with an invariant mass near $m_{\Delta}$. After applying all selections in the example of \tablename.~\ref{tab:cutflow}, the backgrounds are dominated only by the irreducible $\ell^{+}\ell^{-}Z/\gamma$ events.

\section{Significance and exclusion bounds}

\begin{figure}[t]
\centering
\includegraphics[scale=0.98]{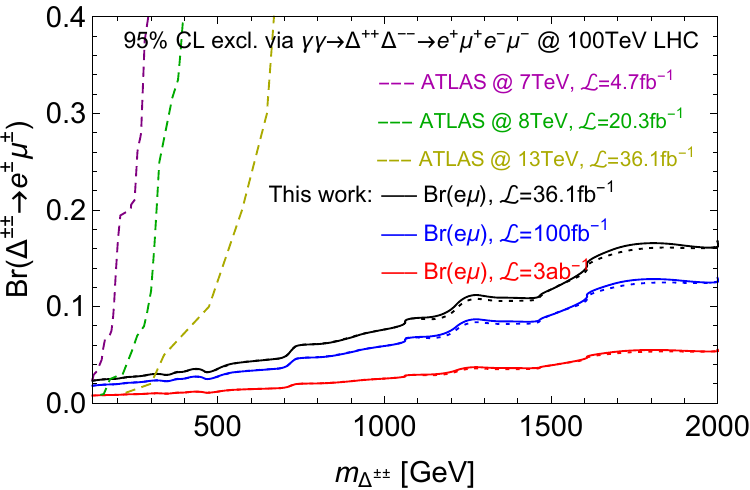}
\caption{95\% C.L. exclusion bounds on the LFV decay branching ratios Br$(\Delta^{\pm\pm}\to e^{\pm}\mu^{\pm})$ versus the triplet scalar mass $m_{\Delta}$ from photon fusion search at the 100 TeV LHC. The results of the present work is displayed in solid (dotted) lines for the normal (inverted) hierarchy of neutrino mass, with black, blue and red colors corresponding to integrated lunimosities of 36.1 fb$^{-1}$, 100 fb$^{-1}$ and 3 ab$^{-1}$. Dashed lines of purple, green and yellow colors are results from the ATLAS experiment~\cite{ATLAS:2012hi,ATLAS:2014kca,ATLAS:2017xqs}.}
\label{fig:exclBr}
\end{figure}

For each scanned value of the scalar mass $m_{\Delta}$, we compute the expected signal significance using the formula:
\begin{align}
\alpha=S/\sqrt{B+(\beta B)^{2}}\,,
\end{align}
where $S$ and $B$ represent the number of signal and background events surviving all selection cuts for a given integrated luminosity, and $\beta$ denotes the systematic uncertainty, which we assume to be 5\%. For three benchmark integrated luminosities: 36.1 fb$^{-1}$, 100 fb$^{-1}$ and 3 ab$^{-1}$, we extract the corresponding $2\sigma$ exclusion limits in the parametric space of the LFV decay branching ratio Br$(\Delta^{\pm\pm}\to e^{\pm}\mu^{\pm})$ versus the scalar mass $m_{\Delta}$, as shown in \figurename.~\ref{fig:exclBr}. Our results are displayed in solid lines, each of which is accompanied by a dotted line with the same color, corresponding to the NH (solid) and IH (dotted) cases, respectively. Black, blue and red colors correspond to luminosities of 36.1 fb$^{-1}$, 100 fb$^{-1}$ and 3 ab$^{-1}$. Experimental results from the ATLAS collaboration are also shown in dashed lines for comparison, where the purple, green and yellow ones correspond to colliding energies of 7, 8 and 13 TeV with luminosities of 4.7, 20.3 and 36 fb$^{-1}$~\cite{ATLAS:2012hi,ATLAS:2014kca,ATLAS:2017xqs}.

It can be seen from the contours that with a sufficient increase of colliding energy and collected data, the sensitivity for the LFV decay branching ratio  Br$(\Delta^{\pm\pm}\to e^{\pm}\mu^{\pm})$ can be improved significantly. In the small mass range of the triplet scalar below $\sim300$ GeV, the searches through Drell-Yan processes exhibit better sensitivity than that of photon fusion in the present work. However, as the triplet mass increases, photon fusion search demonstrates better probing capability, and much smaller branching ratios can be probed or excluded at the TeV scale. Even with a relatively small amount of data 36.1 fb$^{-1}$ (black solid and dotted lines in \figurename.~\ref{fig:exclBr}), the exclusion bound for the branching ratio can reach $\sim10\%$ for as large as 1 TeV of the triplet mass. The probing sensitivity can be improved significantly to Br$(\Delta^{\pm\pm}\to e^{\pm}\mu^{\pm})\approx1\%$ in the TeV region if the luminosity can be increased to 100 fb$^{-1}$ and 3 ab$^{-1}$ (blue and red lines in \figurename.~\ref{fig:exclBr}).

One can also infer from the figure that contours for the NH and IH cases are very similar to each other, as expected, due to the similar kinematic distributions for each case (see \figurename.~\ref{fig:dist}) as discussed in Section.~\ref{sec:search}. But with the best-fit values for neutrino oscillation parameters published by the NuFIT program (Eq.~\eqref{pmns:IH} and Eq.~\eqref{pmns:NH})~\cite{Esteban:2024eli,nufit}, the branching ratio Br$(\Delta^{\pm\pm}\to e^{\pm}\mu^{\pm})\approx2.6\%$ for the IH neutrino mass hierarchy while the ratio approximates to 0.5\% for the NH case. Hence, with these values of the branching ratio, the $2\sigma$ exclusion for the triplet scalar mass reaches $m_{\Delta}\approx1150$ GeV for the IH case, while the mass exclusion for the NH case is only hundreds of GeV. The probing sensitivity for the IH case can reach a larger mass exclusion limit than that from the ATLAS experiment in the parametric space and can exceed the current bound of 1080 GeV given by using 139 fb$^{-1}$ Run-2 data of the LHC~\cite{ATLAS:2022pbd}. It should be noted that pile-up effects are not considered in the present study, which is left for our future study. For one of the effective ways, the readers are suggested to refer to related papers utilizing time-of-flight detectors to suppress the combinatorial background in new physics searches ~\cite{Tasevsky:2014cpa,Cerny:2020rvp,Goncalves:2020saa,Tasevsky:2022sch}.

\section{Conclusion}
In the present paper, we explore the sensitivity of probing the type-II seesaw doubly-charged triplet scalars $\Delta^{\pm\pm}$, followed by the lepton-flavor-violating decay into same-sign $e^{\pm}\mu^{\pm}$ pairs through elastic photon fusion production. A degenerate mass spectrum for $SU(2)_{L}$ triplet scalars and a tiny vacuum expectation value $v_{\Delta}<10^{-4}$ GeV are assumed so that leptonic decay modes are dominant over the bosonic and cascade decay modes. To improve the sensitivity, we consider a high colliding energy of 100 TeV with different scenarios of luminosities at the LHC. With the input of best-fit values from neutrino oscillations for the neutrino mixing matrix, exclusion limits at 95\% C.L. for the triplet mass $m_{\Delta}$ are obtained under both normal and inverted hierarchies. With an integrated luminosity of 3 ab$^{-1}$ and an assumption of inverted hierarchy, the mass exclusion bound can reach around 1150 GeV, surpassing the current constraint at 1080 GeV from Run-2 data.

\section{Acknowledgments}
This work is supported by the National Natural Science Foundation of China under Grant No. 12405118 and the Natural Science Foundation of Jiangsu Province under Grant No. BK20230623.

%=========================================================================================================================
% The bibliography will probably be heavily edited during typesetting.
% We'll parse it and, using the arxiv number or the journal data, will
% query inspire, trying to verify the data (this will probalby spot
% eventual typos) and retrive the document DOI and eventual errata.
% We however suggest to always provide author, title and journal data:
% in short all the informations that clearly identify a document.
%=========================================================================================================================

\end{document}